\documentstyle[12pt]{article}
\def\beq{\begin{equation}}
\def\eeq{\end{equation}}
\def\bear{\begin{eqnarray}}
\def\ear{\end{eqnarray}}
\def\intl{\int\limits}
\def\nn{\nonumber\\ {}}

\begin{document}

\title{ Explanation of the anomalous acceleration of a free moving
body in the frame of extended space model }

   \author{D. Yu. Tsipenyuk\thanks{e-mail:tsip@kapella.gpi.ru} \\
   \normalsize General Physics Institute of the Russian Academy of Sciences \\
   \normalsize 119991, Russia, Moscow, Vavilova str.38 \\
   \normalsize and \\
    W. B. Belayev\thanks{e-mail:dscal@ctinet.ru} \\
   \normalsize Center for Relativity and Astrophysics,\\
   \normalsize 185 Box , 194358, Sanct-Petersburg, Russia}

\maketitle

\begin{abstract}

We consider (1+4) generalization of classical electrodynamics
including gravitation field. With this approach it is assumed a
presence of an extra component of extended field stress tensor,
whose physical interpretation is based on necessity to obtain
Newton gravity law as particular case.

In model analyzed in present work mass of having in 4D a rest mass
particle moves in five-dimensional space-time along fifth
coordinate with light velocity and its electric charge is a
stationary value in additional dimension. Generalized gravity law
that is obtained from extended Maxwell equations confirmed by
existence of variable component of Pioneer 10 additional
acceleration, whose analyze is made in frame of this model.
\end{abstract}

Generalization of the special theory of relativity (SR) for
5-dimensional extended space with the metric (+; -, -, -, -) has
been offered in the papers [1-11]. In particular, this model of
extended space (ESM) allows integrating electromagnetic and
gravitational interactions.

In ESM interval $S$ (which already exists in the (1 + 3)
-dimensional Minkowski space $M(T;\vec{X})$ ) is used as the 5-th
additional coordinate of the (1 + 4) -dimensional extended space
$G(T;\vec{X};S)$:

 \beq
 S^{2}=(ct)^{2}-x^{2}-y^{2}-z^{2}.
 \eeq

Interval $S$ is conserved at common Lorentz transformations in the
Minkowski space $M(T;\vec{X})$ but varies at turns in the extended
space $G(T;\vec{X};S)$. Thus, Minkowski space $M(T;\vec{X})$ is a
cone in extended space $G(T;\vec{X};S)$. A stable particle, which
has a rest mass in the 4-D Minkowski space, has not a rest mass in
5-D space independence from turns in this 5-D space. This particle
moves along fifth coordinate with speed of light (in case of zero
5-interval). Such approach corresponds to the Kaluza-Klein theory
\cite{12} and theory of induced substance \cite{13}.

Other version will be existence of a particle in 5-D space, which
can be stationary in this space and has a rest mass (nonzero
5-interval). Not limiting a generality, we will consider version
with a zero 5-interval.

The various details of the ESM structure were considered in papers
[4-10], and also repeatedly were reported on various scientific
conferences \cite{11,14,16,17}. The main difference ESM from SR is
that ESM allows to study processes at which mass m of a particle
is an alternative value. In the ESM model mass of a particle is a
component of 5-vector in space $G(T;\vec{X};S)$ other components
are energy and momentum of a particle. Thus, mass of a particle
varies at transformations in $G(T;\vec{X};S)$.

In articles \cite{1,8} (see also \cite{4,16}) in the ESM were
constructed:

5-vector potential -
 \beq
 (\varphi,\vec{A},A_4)=(A_T,A_x,A_y,A_z,A_S),
 \eeq
the stress 5x5 tensors -
 \beq
 F_{ik}=\frac{\partial A_i}{\partial x_k}-\frac{\partial A_k}{\partial x_i}
 \eeq
and energy-momentum-mass tensors -
 \beq
 T^{ik}=\frac{1}{4\pi}(-F^{il}F^{k}_l+\frac{1}{4}g^{il}F_{lm}F^{lm});\
 i,k=0,1,2,3,4
 \eeq
Here $\|g^{ik}\|$ - metric tensor of the extended space. The
stress tensor was recorded as:
 \beq\label{f5}
 \| F_{ik}\|=\left(\begin{array}{ccccc}
 0&-E_x&-E_y&-E_z&-Q\\
 E_x&0&-H_z&H_y&-G_x\\
 E_y&H_z&0&-H_x&-G_y\\
 E_z&-H_y&H_x&0&-G_z\\
 Q&G_x&G_y&G_z&0\\
 \end{array}
\right) \
 ,
 \eeq
where the elements with indexes $i, k = 0,1,2,3$ correspond to
components of stress tensor of an electromagnetic field in 4-D
space-time. Vector $\vec{G}=(G_x,G_y,G_z)$ and scalar $Q$ are
additional components.

In the element of energy-momentum-mass tensor, which associated
with density of energy, are added additional members $G^2$ and
$Q^2$:
 \beq
 T^{00}=\frac{1}{8\pi}(E^2+H^2+G^2+Q^2).
 \eeq
Modified vector of the momentum density looks as:
 \beq
 \vec{P}=\frac{1}{c}(T^{01}+T^{02}+T^{03})=
 \frac{1}{4\pi c}([\vec{E},\vec{H}]-Q\cdot \vec{G}).
 \eeq
Poynting vector of a field is expressed through vector of a
momentum as  $\vec{S}=c^2\vec{P}$ .

Vector of the current constructed in the paper \cite{8}:
 \beq
 \vec{\rho}=(\rho,\vec{j},j_s),
 \eeq
where $\rho, \vec{j}$ are densities of electric charge and
current, $j_s$ is additional component, corresponds to coordinate
$s$.

In present paper we will turn to account formalism allowing
separation electric charge from mass. It consists in assumption
that if massive particle moves in fifth dimension with light
velocity his electric charge is stationary in this dimension. But
in 4D mass and electric charge of the particle have the same
coordinates. Introduction of this formalism is conditioned, first,
by necessity of fulfillment of the Maxwell equations and,
secondly, by necessity of an absence of electric current in the
open form in the given below equation (\ref{f15}), defined
gravitation law, since otherwise presence of stationary in 4D
electric charge influences on gravity. Such assumption does not
contradict to the Kaluza-Klein theory and may be considered as
version of ESM.

By means of the stress tensor (\ref{f5}) and the vector of the
current extended Maxwell equations have been obtained in \cite{8}:
 \bear
   div\vec{H}=0, \\
   rot\vec{E}+\frac{1}{c}\frac{\partial \vec{H}}{\partial t}=0, \\
   rot\vec{G}+\frac{\partial \vec{H}}{\partial s}=0, \\
   \frac{\partial \vec{E}}{\partial s}+\frac{1}{c}\frac{\partial \vec{H}}{\partial t}+gradQ=0,
   \\
   div\vec{E}+\frac{\partial \vec{Q}}{\partial s}=4\pi\rho, \\
   rot\vec{H}-\frac{\partial \vec{G}}{\partial s}-\frac{1}{c}\frac{\partial \vec{E}}{\partial
   t}=\frac{4\pi}{c}\vec{j}, \\
   \label{f15}
   div\vec{G}+\frac{1}{c}\frac{\partial \vec{Q}}{\partial s}=4\pi j_s.
 \ear
Vector $\vec{G}$ in \cite{4} is related to gravitation field and
it is necessary to note, that in the equation (\ref{f15}) the only
mass current $j_s$ is included, since otherwise a stationary
electric charge in 4D would influence on gravity. As components of
the stress tensor have to be identical we set vector $\vec{G}$ as
:
 \beq\label{f16}
 \vec{G}=q\vec{V},
 \eeq
where $\vec{V}$ is the stress vector of the gravitation field, $q$
is a constant having dimension $[q] = [{\em mass\/}]/[{\em
electric charge\/}]$.

In the present work we will determine the field $Q$ by the
following expression: \beq \label{f17}
 Q=4\pi c\gamma q\intl_{t_0}^{t}\mu(t,x,y,z,s)dt,
\eeq
 where $\gamma$ is gravitational constant, $\mu$ is mass
density, $t_0$ is a constant of time. Introduction of field $Q$ in
this generalize form is conditioned by the demand that Newton
gravity has to be a particular case with substitution one in
equation (\ref{f15}). In this interpretation field $Q$ bears a
relation to mass but this component of stress tensor has not a
direct connection with mass, since we take derivative of $Q$ with
respect to time.

Current density $j_s$ we assume to be null, which describes
absence of the electric charges movement along the fifth
coordinate. Let us consider example of field when electrical and
magnetic components of the stress tensor are default, current
vector and potentials are the following:
 \bear
 \vec{\rho}=0, \ \varphi=0, \ \vec{A}=0 \nn
 A_s=qW+2\pi c\gamma q\int\intl_{\tau_0}^{\tau (t)}\mu (x,y,z)d\tau (t)dt+C,
 \ear
where $W$ is a gravitational potential that expressed through
vector $\vec{V}$ as: $V_i=\frac{\partial W}{\partial x_i},
i=1,2,3$, $\tau$ is a proper time in the point $X=(x,y,z)$ ,
$\tau_0$ and $C$ are constant. We take $W$ to be independent from
time and fifth coordinate $s$. Introduction of coordinate time in
point $X$ is conditioned by the fact that field $Q$ (\ref{f17}) is
depended from time, which of course may differentiate in points
$X_0$ and $X$ \cite{18}. Here point $X_0$ sets down as a center of
the coordinate system.

Thus the system of extended Maxwell equations gives only
non-vanish equation (\ref{f15}), which taking into account
(\ref{f16}) turns out to be
 \beq\label{f19}
 div\vec{V}+4\pi\gamma\frac{\partial\tau}{\partial t}\mu(x,y,z)=0.
 \eeq
With $\tau=t$ that means an equality of proper and coordinate
time, this equation is transformed to the well-known Poisson
equation for gravitation field.

Let us employ this approach for the analysis of the
Pioneer-effect. NASA launched Pioneer 10 on March 1972 towards
distant planets of the solar system. Radiometric data shows that
additional acceleration takes place in direction of the Sun
\cite{19,20,21}. In \cite{22} the constant component of this
acceleration is considered as manifestation of non-orthogonal
metric of five-dimensional space-time.

As was noticed earlier we don't consider a case when massive
particle in 5D is stationary in space and has a rest mass, and its
motion consists with non-zero 5D interval:
 \beq\label{f20}
 (ct)^2-x^2-y^2-z^2-s^2=const>0.
 \eeq
However, we note that in gravitational field the photon velocity
is less then light velocity and this mathematically gives a
nonzero value of the 5D interval in (\ref{f20}) for describing the
movement in this field \cite{9,10}. Then for explanation of
constant component of Pioneer 10 anomalous acceleration existence
in framework of ESM the approach described in \cite{22} is
allowed, since non-zero 5D interval is considered for metric
introduced here. This possibility will be a subject of our further
research. Review of theories that explained an existence of the
additional acceleration is contained in \cite{20,21}. Taking into
account type of the Pioneer 10 mission it is possible to consider
as trustworthy the data from 1987 to 1999 years \cite{20}, when
its distance to the Sun was increased from $40$ to $60$ ${\rm
AU}$. In work \cite{23} variable component observed additional
acceleration is related with periodic change of a velocity of
Pioneer 10 in relation to the Earth, which is caused by its
movement around the Sun. Velocity of the Pioneer 10 in relation to
the Earth $v$ approximately determined by the expression
 \beq\label{f21}
 u=u_P+u_E cos(\nu t+\varphi_0),
 \eeq
where $u=v/c$, $u_P\approx 4.13\times 10^{-5}$ corresponds to the
Pioneer 10 velocity in relation to the Sun and $u_E=9.97\times
10^{-5}$ corresponds to the orbital Earth velocity, $\gamma=2\pi$
${\rm 1/year}$, $\varphi_0$ is a constant.

Thus the Pioneer 10 proper time in Earth's center coordinate
system neglecting terms containing $u$ of higher orders of
magnitude will be
 \beq
 \tau = (1-u)t.
 \eeq
Taking in account this relation the equation (\ref{f19}) has a
solution in a form of a potential in point $(x_P,y_P,Z_P)$:
 \beq
 W=(1-u)\gamma\int{\frac{\mu
 d\Omega}{\sqrt{(x-x_P)^2+(y-y_P)^2+(z-z_P)^2}}},
  \eeq
where $d\Omega$ is volume element. This potential can be rewritten
as:
 \beq\label{f24}
 W=(1-u)\gamma\frac{M}{r},
 \eeq
where $M$ is Sun's mass, $r$ is a distance from the Sun to the
Pioneer 10. By comparison with Newton gravity obtained potential
gives additional acceleration
 \beq\label{f25}
 \hat{a}=u\gamma\frac{M}{r^2},
 \eeq
to the massive particle with a radial movement.

It may be made still one supplementary prediction following from
ESM on additional acceleration originated from $u_P=4.13\times
10^{-5}$ taking into account (\ref{f21}) and (\ref{f24}). But its
value will be smaller by an order than the full acceleration
(\ref{f25}).

We obtain from the equation (\ref{f25}) that amplitude of the
oscillation of periodic component of the velocity (\ref{f21})
$u_E$ has the amplitude of additional variable acceleration
$|{a^{E}_P}|=3.7\times 10^{-8}$ ${\rm cm/s}^2$ at the distance
$r=40$ ${\rm AU}$ and $|{a^{E}_P}|=1.6\times 10^{-8}$ ${\rm
cm/s}^2$ at $r=60$ ${\rm AU}$. For the Pioneer 10 this amplitude
determined from measurements \cite{20} amounts correspondingly to
$|{a^{E}_P}|_{40AU}=(2.9-2.4)\times 10^{-8}$ ${\rm cm/s}^2$ and
$|{a^{E}_P}|_{60AU}=(1.3-0.8)\times 10^{-8}$ ${\rm cm/s}^2$.

Thus, we considered (1+4) generalization of classical
electrodynamics including gravitation field. With this approach it
is assumed a presence of an extra component of extended field
stress tensor, whose physical interpretation is based on necessity
to obtain Newton gravity law as particular case.

In this work we have analyzed model where mass of having in 4D a
rest mass particle moves in five-dimensional space-time along
fifth coordinate with light velocity and its electric charge is a
stationary value in additional dimension. Generalized gravity law
that was obtained from extended Maxwell equations confirmed by
existence of variable component of Pioneer 10 additional
acceleration, whose analyze was made in frame of this model.

\end{document}